\documentstyle[12pt]{article}
\begin{document}
\pagenumbering{arabic}
\title{Interaction of Vortices with an External Field}
\author{Ya.~Shnir
\\[2mm]
{\it  {\normalsize Institut f\"ur Physik, Universit\"at Oldenburg
}}\\
{\it  {\normalsize D-26111 Oldenburg, Germany}}
\date{~}
}
\maketitle
\begin{abstract}
The interaction of a magnetic flux vortex  with weak external fields
is considered in the framework of the Abelian Higgs model. 
The approach is based on the calculation of the zero-mode excitation 
probability in the external field. The excitation of the field configuration 
is found perturbatively. As an example we consider the effect of 
interaction  with an external current. The linear in the scalar field 
perturbation is also considered.
\end{abstract}

\maketitle

\section*{Introduction}

As is known, the Abelian Higgs model in 2+1 dimensions possesses a finite-energy 
static 
topological nontrivial 
solution of the vortex type \cite{NielsOles}. In the nonrelativistic limit the
model has the form of the Ginzburg-Landau theory, which phenomenologically 
describes the magnetic flux 
vortices in the type II superconductor \cite{Abrikos}. On the other hand, the 
extrapolation of the vortex solution of the Abelian Higgs model to 3+1 dimensions
gives $U(1)$ string configuration, which could be produced as a topological defect
at the early Universe \cite{Kibble}. Very recently a new class of closed vortex 
ring solutions was discovered in  3+1 dimensional $SU(2)$ Yang-Mills-Higgs theory
\cite{KKS}.  
Thus the strings or vortices have played many
interesting roles in the interplay between high energy physics, condensed matter and 
cosmology.  

The dynamical properties of vortices in the
Abelian Higgs model and some of its modifications 
have been considered from different points of view.
The much studied question seems to be a problem of interaction between strings.

The general proof of the existence of multivortex configurations was 
constructed in \cite{Weinberg,Taubes} 
and a 
method for obtaining asymptotic multisoliton solutions in gauge theories was 
given
in  \cite{Ball}. There is a nice description of vortices dynamics based on the
moduli space technique \cite{Rubak,Samols}. In this scheme low energy soliton
dynamics is approximated by geodesic motion on the space of the collective 
coordinates of static multivortices configuration with respect to the metric
induced by functional of the kinetic energy. Since no exact 
static multivortices solutions are known, some numerical calculations have been
performed.
The numerical investigation of the interaction between well-separated 
vortices was carried out by Shellard and Ruback \cite{Shell}. 
An analytical study
of the interaction between vortices was done recently in \cite{Betten}.

The general result of these works is that there are no long-range forces 
between static vortices. There is a difference from BPS multimonopole
configuration where the repulsive and attractive long-range forces
between monopoles exactly compencate, BPS monopoles experience no net
interaction. That makes the problem of the string dynamics in an
external field more compicated comparing to the case of
monopoles. Actially the interaction of vortices with external fields was
not much studied. 
  
There is a regular perturbation scheme used to describe
the motion of solitons under external force \cite{KisSh}.
This approach is based on the calculation of the probability of the 
excitation  of
corresponding translations zero modes in an external field. In our previous
publication \cite{KisSh} we  
discussed the 
application to the problem of motion of 1+1 dimensional kink in 
$\phi^4$ model and to the case of the interaction between the 
't Hooft-Polyakov monopole and an 
external weak field.

In the present note we would like to apply 
this formalism to the problem of 
interaction between 2+1 dimensional vortex solution in 
the  Abelian Higgs model and an external field.  

\section{Current-current interaction}

The Lagrangian of the Abelian Higgs model in (2+1) dimensional space-time 
is given 
(in the units $c=1, ~\hbar=1$):
\begin{equation}                 \label{Lagrang-Higgs}  
 L =  \frac{1}{4} {F_{\mu \nu}} {F^{\mu \nu}} + \frac {1}{2} (D_{\mu}\phi)^*
(D^{\mu}\phi )  + \frac {\lambda}{4}
\left(|\phi |^2 - v^2\right)^2,
\end{equation}
where a $U(1)$ real gauge potential $A_\mu = (A_0, A_k)$,  
$k = 1,2$,  is coupled to a charged complex scalar field $\phi = \phi_1 + 
i\phi_2$.
Here $e$ is the gauge coupling constant, 
$\lambda$ is the scalar field self-coupling constant, 
$v$ is the vacuum expectation
value of the modulus of the scalar field  and 
$$      
F_{\mu \nu} = \partial _{\mu} A_{\nu} -  \partial _{\nu} A_{\mu}; 
$$
\begin{equation}
\quad 
D_{\mu}\phi = \partial _{\mu} \phi  - ie A_{\mu}\phi ,  ~~{\rm or }~~
D_{\mu}\phi^a = \partial _{\mu} \phi^a  - e\varepsilon_{ab}
A_{\mu}\phi^b,~~~~ a = 1,2
\end{equation}
are the field strength tensor and the covariant derivative correspondingly.  
Hereafter, the space indices are $m,n,k \dots = 1,2$ and the 
real components of the complex 
Higgs field are labeled by the indices $a,b,c \dots = 1,2$.

This model is a relativistic analog of the Landau-Ginzburg theory of 
syperconductivity.  
The finiteness of the energy determines the asymptotic form of the fields.
There are  nontrivial static vortex solutions of this model
depending on the polar coordinates $r$ and $ \varphi$ \cite{NielsOles}, 
\cite{Abrikos} which are described by  
the Nielsen-Olesen ansatz:  
\begin{equation}                  \label{Anz-Niel-Olesen}    
A_0  =  A_r = 0;~  A_{\varphi} =   \frac{n}{er}\left(1 - K(r)\right) ;\quad
\phi  = v H (r) e^{in\varphi}, ~~{\rm where}~ n \in Z. 
\end{equation} 
Here the boundary conditions on the structure functions $K(r)$, $H(r)$ are:
$$
K(r)~\buildrel r\to\infty\over \longrightarrow 0, \qquad 
H(r)~\buildrel r\to\infty\over\longrightarrow 1;
$$
$$
K(r)~\buildrel r\to 0\over \longrightarrow ~1, \qquad  H(r) 
~\buildrel r\to 0\over \longrightarrow~0
$$ 
and the only nonzero component of the field strength tensor is $F_{xy} = 
-(n/er) K'(r) = B_z$.

Let us consider the interaction of this configuration with an external
field. Note that because both vector and scalar fields are coupled in 
the vortex configuration,
there are a few different ways to introduce such an interaction. 
For example, one could add to the Lagrangian 
(\ref{Lagrang-Higgs}) an  extra current-current perturbation term that is 
linear in external current $J_m$  as well as in current of scalar field
 \begin{equation}                                   \label{Lagrang-interact-v}
L_{int}^{(v)} = j_{m} J_{m} \equiv \frac{ie}{2} 
\left[\phi^* (D_{m }\phi) - (D_{m }\phi)^* \phi \right] J_{m}.
\end{equation}

As a result the classical field equations take the form
\begin{eqnarray}                              \label{field-eq-Higgs}
\partial _{m} F_{m n} &= & \displaystyle 
\frac {ie}{2} \left[\phi^* (D_{n }\phi) - (D_{n }\phi)^*  \phi \right]
+ {\cal F}_{n}^{(1)};
\nonumber  \\
D_{m }D_{m } \phi^a  &= & \lambda (\phi^b \phi^b - v^2) \phi^a + 
{{\cal F}^a}^{(2)}
\end{eqnarray} 
where the last terms represents the external force acting on the 
configuration. They read

\begin{eqnarray}     \label{force}
{\cal F}_{n}^{(1)} = e^2 |\phi|^2 J_{n};
\qquad {{\cal F}^a}^{(2)} = 2e~ \varepsilon_{ab}
D_{n } \phi^b J_{n}  
\end{eqnarray}

The effect of such a perturbation is that there are corrections to the static 
vortex solution (\ref{Anz-Niel-Olesen}). In an analogy with the case of
two-dimensional $\lambda\phi^4$ model \cite{KisSh} these 
corrections could be expanded in powers of external perturbation:
$$  
A_{m} = (A_{m})_0 +  a_{m} + \dots;\qquad 
\phi ^a = (\phi ^a)_0 +  \chi^a + \dots 
$$ 
where $(A_{m})_0, (\phi ^a)_0$ correspond to the classical
$n$-vortices solution given by the zeroth-order approximation 
(\ref{Anz-Niel-Olesen}) and the fluctuations of the vector and scalar fields  
on this background are of the same order as perturbation $J_m$.
 
To the first order corrections, they can be found from the 
equations describing the fluctuations of vector and scalar fields:
\begin{eqnarray}                             \label{second-order-vortex}
\left(-\frac{d^2}{dt^2} + \partial _{m} \partial _{m} - e^2\phi^a \phi^a
\right)  a_{n} &=&-2e \varepsilon_{ab}D_{n}\phi^a \chi^b + {\cal F}_{n}^{(1)}; 
\nonumber\\
\left(-\frac{d^2}{dt^2} + D_{m} D_{m}\right)\chi^a&=& 
2e\varepsilon_{ab}D_{m}{\phi}^b a_{m} + {{\cal F}^a}^{(2)}, 
\end{eqnarray}
where we have decomposed the complex field via $\chi = \chi_1 + i\chi_2$,
and the background field gauge $\partial_{m} a_{m}
 = ie (\phi ^* \chi- \chi ^* \phi)$ is used. 

Now  one can  apply the same approach that was already used in  \cite{KisSh}, i.e.
write the expansion of the fields $
a_{m}({\bf r},t)$, $\chi^a ({\bf r},t)$ on the eigenmodes of the
matrix ${\cal D}^2$
of second functional derivatives of the action 
with respect to the fields $A_{m}, \phi^a$ 
$$ 
{\cal D}^2 \left(\begin{array}{c} 
a_{n}\\ 
\chi ^a\end{array} \right) \equiv \left( 
\begin{array}{cc} 
(\partial_{m}^2 - e^2 \phi^a \phi^a)a_n   & 2 e \varepsilon_{ab} 
D_{n} \phi^a \chi^b \\[3pt] 
-2 e \varepsilon_{ab} D_m  \phi^b a_m  & D_m D_m \chi^a 
\end{array}  \right)  . 
$$ 
In matrix notation the equation of motion (\ref{second-order-vortex}) 
can be rewritten in the form
\begin{equation}      \label{second-matrix}
\left(-\frac{d^2}{dt^2} + {\cal D}^2 \right) f = {{\cal F}},
\end{equation} 
where 
$$
 f =  \left(\begin{array}{c} 
a_{n}\\ 
\chi ^a\end{array} \right);\qquad {{\cal F}} = \left(\begin{array}{c} 
{\cal F}_{n}^{(1)}\\ 
{{\cal F}^a}^{(2)}\end{array} \right).
$$ 

We seek for the solution of Eq. (\ref{second-matrix}) in the form of 
an expansion 
\begin{equation}                                              \label{fexpan} 
f ({\bf r}, t)=\sum\limits_{i=0}^{\infty}C_i(t)\zeta_i({\bf r}) 
\end{equation} 
on the complete set of eigenfunctions $\zeta_i({\bf r})$ of the operator 
${\cal D}^2$. These eigenfunctions consist of a vector and a scalar 
component: 
$\zeta_i({\bf r}) = \left(\begin{array}{c} 
\eta_{mi}({\bf r})\\ 
\eta^a_i({\bf r})\end{array} \right)$ 
describing the fluctuations of the corresponding fields on the vortex
background \cite{Weinberg}, \cite{Rubak}. Thus, there are indices of two
kind: index $i$ is the number of a 
mode and the indices $m$,$a$ correspond to its
spatial and `isotopic' components. 

The substitution of 
expansion (\ref{fexpan}) into Eq.(\ref{second-order-vortex}) 
results in the following system of equations for coefficients $C_i(t)$: 
\begin{eqnarray}                            \label{expan-YM-modes} 
\sum _{i = 0}^{\infty} \left({\ddot C}_i + \Omega_i^2 C_i \right) 
{\eta_{mi}({\bf r})} - 2 e \varepsilon_{ab} \chi^b D_{m} \phi ^a 
=   {{\cal F}_{m}^{(1)}({\bf r})}; 
\nonumber\\ 
\sum _{i = 0}^{\infty} \left({\ddot C}_i + \omega_i^2 C_i \right) 
{\eta ^a_i ({\bf r})} - 2 e \varepsilon_{ab} a_{m} D_{m} \phi ^b 
=  {{\cal F}^a}^{(2)}({\bf r}), 
\end{eqnarray} 
This is a system describing two sets of coupled forced oscillators.

It is known that among all fluctuations on the vortex background there 
are modes with zero energy ($\Omega_0 = \omega_0 = 0$) \cite{Weinberg}.
Such modes are collective translation coordinates  
of the multivortices configuration. All the other solutions of the system
(\ref{expan-YM-modes}) correspond to the oscillations on the classical
background. Unfortunately neither analitical solution for the vortex 
structure functions nor the eigenfunctions ${\eta_{mi}({\bf r})},
{\eta ^a_i ({\bf r})}$ are known. Nevertheless, some information about
the vortex dynamic can be obtained from the system  
(\ref{expan-YM-modes}).
 
Following the approach by Manton \cite{Manton-77}, one can find 
the acceleration
of the vortex under perturbation $J_m$ if 
the excitations of the zero modes are treated 
as a nontrivial time dependent
translation of the n-vortex configuration (\ref{Anz-Niel-Olesen}). 

However, the structure of the Lagrangian (\ref{Lagrang-Higgs}) 
suggests \cite{Rossi} 
that the   normalizable zero modes are not only translations of the   
topologically non-trivial configurations but  
$${\zeta}^{(k)}({\bf r}) \equiv \zeta_0({\bf r})= \left(\begin{array}{c} 
{{\eta_{m}({\bf r})}}^{(k)}\\ 
{{\eta}^a({\bf r})}^{(k)}\end{array} \right)$$ 
where 
\begin{eqnarray}                   \label{modeYM-zero} 
{ \eta_n({\bf r})}^{(k)} = F_{kn} = 
\partial _k  A_n - \partial_n A_k; \quad 
{\eta ^a({\bf r})}^{(k)} = D_k \phi^a = \partial _k \phi ^a - e 
\varepsilon_{ab} \phi^b A_k. 
\end{eqnarray} 
Here the index $k$ corresponds to the translation  in the direction ${\hat 
r}_k$. 

Similarly to the case of 3+1 dimensional Georgi-Glashow model, 
these normalizable zero modes coincide with the pure 
translational quasi-zero 
modes of the vector and scalar fields ~ ${{\widetilde \eta_n} ({\bf 
r})}^{(k)} = \partial _k  A_n; ~ {{\widetilde \eta^a}({\bf r}) }^{(k)} = 
\partial _k \phi ^a $  up to a gauge transformation with a special choice 
of the parameter which is just the gauge potential $A_k$ itself. 
These modes are normalized in such a way that makes $C_0$ in 
expansion (\ref{fexpan}) equal to the displacement of the vortex in the 
direction ${\hat r}_k$:
\begin{eqnarray}   
A_n({\bf r}+\delta {\bf r}) \approx A_n({\bf r}) + 
\partial_k A_n({\bf r}) \delta x_k =
A_n({\bf r}) + C_0(t){{\widetilde \eta_n}} ({\bf r})}^{(k);\nonumber\\
\phi^a({\bf r}+\delta {\bf r}) \approx  
\phi^a({\bf r}) + \partial_k \phi^a({\bf r}) \delta x_k =
\phi^a({\bf r}) + C_0(t){{\widetilde \eta^a}}({\bf r})^{(k)}
\end{eqnarray}

Now we can project the Eq.(\ref{expan-YM-modes}) onto 
the zero modes (\ref{modeYM-zero}) which yield the equation: 
\begin{eqnarray}                              \label{modeYM} 
&&{\ddot C}_0 \int d^2x \left[\left({\eta^a({\bf r})}\right)^2 + \left(
{\eta_m({\bf r})}\right)^2\right] - 2e \int d^2x~
\varepsilon_{ab}\left\{D_m\phi^a \chi^b{\eta_m} + a_m D_m \phi^b {\eta^a}
\right\}
\nonumber\\
&=&\int d^2 x {{\cal F}_m({\bf r})} 
{{\eta _m ({\bf r})}}^{(k)} + \int d^2 x  {{\cal F}^a({\bf r})} 
{\eta ^a({\bf r})}^{(k)}. 
\end{eqnarray} 

The second term on the left-hand side of this equation 
describes the transitions between the vortex zero modes ${\eta^a}, {\eta_m}$ 
and all the 
other fluctuations $\eta^a_i, \eta_{mi}$ on the vortex background.
A substitution of the expansion  (\ref{fexpan}) gives 
\begin{eqnarray}         \label{trans}
{\phantom{=}} &-&2e \int d^2x~
\varepsilon_{ab}\left\{D_m\phi^a \chi^b{\eta_m} + a_m D_m \phi^b {\eta^a}
\right\}\nonumber\\ 
 = &-&2e \int d^2x~
\varepsilon_{ab}\sum _{i = 0}^{\infty} C_i(t) \left\{ 
{\eta^a}^{(m)}
{\eta^b}_i{\eta_m} + {\eta^a} {\eta^b}^{(m)}{\eta_m}^i\right\}
\end{eqnarray}

The first term in the sum (\ref{trans}) is equal to zero: the vortex 
configuration behave as a particle-like object and there is no effect of 
mutual transitions between the collective coordinates of scalar and vector 
fields. All the other terms ($i\ne 0$) describe
the effect of a bremsstrahlung of both vector and scalar massive
fields from a vortex accelerated by an external force. 
Such  effects of radiation are suppressed as $\sim \exp\{-m_{s(v)}r\}$ 
and, if  
scalar (vector) fields fluctuation are very heavy, they can be  considered
as an additional small perturbation of the second order. But if $m_{s(v)}$ are
small, the contribution of this transition term (\ref{trans})
could be of the same order as the
probability of the zero mode excitation. We will not 
consider here this situation
and neglect the contribution from the term (\ref{trans}).   

Note, that the first integral on the left-hand side of the Eq.(\ref{modeYM})
gives a very simple result. Because of the virial theorem (see 
e.g, \cite{Radj}), the kinetic energy of a vortex is equal 
to the potential energy and we obtain
\begin{eqnarray}
\int d^2x \left[\left({\eta^a({\bf r})}\right)^2 + \left(
{\eta_m({\bf r})}\right)^2\right] = 
\int d^2x \left(F_{kn}^2 + (D_k\phi)^2\right) \nonumber\\
= \int d^2x \left\{\frac{1}{2}\left(F_{kn}^2 + (D_k\phi)^2 
\right) + V[\phi]\right\} = M
\end{eqnarray}
that is simple the energy of static vortex per unit length, or its mass. 
Note that the same relation between the normalization factors of monopole
zero modes and the monopole mass holds in 't Hooft-Polyakov model 
\cite{KisSh}.

Now we turn to the right-hand side of the Eq.(\ref{modeYM}). 
Suppose that the external constant current $J_m$ is directed along the $x$ axis:
$J_m = (J,0)$.  
Taking into account the definition (\ref{force}) of external force acting 
on a vortex and substitute the ansatz (\ref{Anz-Niel-Olesen}), 
one could find

\begin{eqnarray}                       \label{force-proj}
I_1 = \int d^2 x {{\cal F}_m^{(1)}({\bf r})} 
{\eta _m ({\bf r})}^{(k)}
 &=& e^2 J_m\int dx dy~\phi^a\phi^a F_{km}\nonumber\\ 
I_2 = \int d^2 x  {{\cal F}^a}^{(2)}({\bf r}) 
{\eta ^a({\bf r})}^{(k)} &=& 
- 2e J_m \int dx dy ~\varepsilon_{ab}
D_m \phi^a D_k \phi^b .
\end{eqnarray}
Thus the non-trivial result gives the projection onto the zero mode 
component
${\zeta}^{(y)}({\bf r})$:
\begin{eqnarray}                     
I_1&=& e^2 J\int dx dy~\phi^a\phi^a F_{xy}
= 2\pi e v^2 Jn\int dr H^2(r) \frac{dK(r)}{dr};\nonumber\\ 
I_2 &=&
- 2e J \int dx dy ~\varepsilon_{ab}
D_x \phi^a D_y \phi^b
= -4\pi ev^2 Jn \int dr H(r) \frac{dH(r)}{dr}\nonumber\\
&=&-2 \pi e v^2 Jn,
\end{eqnarray}
i.e. the only zero modes orthogonal to the external force are 
excited{\footnote{It is interesting to compare this conclusion 
with another problem of the interaction between well separated 
vortices discussed in \cite{Sp}, where the intervortex forces
lead to the well known effect of $\pi/2$-scattering 
of two vortices by head-on collision \cite{Rubak}.}}

Obviously, the result depends upon the relation between the masses
of scalar $(m_s^2 = 2 \lambda v^2)$ 
and vector $(m_v^2 = e^2 v^2)$ particles. For example, in the London limit 
$m_s \gg m_v$, and on the distance ranges at $m_s^{-1} \ll r \ll m_v^{-1}$ 
one could neglect
the core structure of scalar field, i.e. suppose that $H \sim 1$ everywhere. 
Then
we have for the second integral in (\ref{force-proj})
$$
2\pi e v^2 Jn \int dr H^2(r) \frac{dK(r)}{dr} \sim
2 \pi e v^2 Jn \int dr \frac{dK(r)}{dr} = 2 \pi e v^2 Jn.
$$ 

Thus, 
taking into account the definition of the magnetic flux $\Phi_n = 
\int d^2x F_{xy} = \frac{2\pi n}{e}$, we finally obtain
\begin{equation}
M{\ddot C}_0 = - 4\pi ev^2 Jn  = - 2 v^2 e^2 J \Phi_n = - 2 m_v^2 J \Phi_n 
\end{equation}
and the acceleration of a vortex along the direction orthogonal to the 
external force is  
\begin{equation}         \label{accel}
W =  \frac{2 J m_v^2}{M} \Phi_n
\end{equation}

\section{A linear perturbation of the scalar field}

Alongside with the Lagrangian of interaction (\ref{Lagrang-interact-v})
there is another possibility to introduce linear on scalar field 
interaction between the vortex scalar field and and external perturbation: 

\begin{equation}                                   \label{Lagrang-interact-s}
L_{int}^s = \varepsilon \frac{v}{2}\left(\phi e^{-im\varphi} + 
\phi^*e^{im\varphi}\right) = \varepsilon v \varepsilon_{ab}\phi^a n^b =
\varepsilon v^2 H(r) \cos(n-m)\varphi,
\end{equation}
where $\varepsilon \ll 1$ is a perturbation parameter and 
$n^a = (-\sin m\varphi, \cos m\varphi)$ is a unit vector. 
Obviously this is
a perturbation of the vortex configuration 
connected with the external scalar field of another vortex of topological charge $m$. 
The case $m=0$ corresponds to the 
topologically trivial external constant scalar field 
coupled with only one component of the vortex scalar field $\phi^2 = v H(r)
\cos \varphi$.

We will see that the effect of such an 
interaction term depends from the topology of external configuration. Indeed,
if both vortices have the same magnetic flux (i.e., if $m=n$), 
the only effect of the
additional term  (\ref{Lagrang-interact-s}) is a small increasement of the 
configuration mass. But if, for example $n-m =1$, this term lifts degeneration 
of vacuum as it takes place in the case of 2D $\lambda \phi^4$ model or
`t Hooft-Polyakov monopole \cite{KisSh}. 

The difference from the above discussed situation 
is that the interaction term
(\ref{Lagrang-interact-v}) now only affects the scalar component of 
the vortex  configuration. Indeed, the field equations are still given by 
Eq.(\ref{modeYM}), but  the external force acting on the vortex is now
\begin{equation}     \label{force-2}
{\cal F}_{m}^{(1)} = 0;\qquad {{\cal F}^a}^{(2)} = 
\varepsilon v ~\varepsilon_{ab}n^b  
\end{equation}
Projection of this formula onto zero modes along $x$-axis gives
\begin{eqnarray}
&{\phantom{=}}&\int d^2 x  {{\cal F}^a({\bf r})} 
{\eta ^a({\bf r})}^{(k)} =\varepsilon v  \int d^2 x  \varepsilon_{ab}
D_x\phi^a n^b \\
&=& \varepsilon v^2 \int d^2 x \left\{H'\cos\varphi \cos(n-m)\varphi +
\frac{n}{r}HK\sin\varphi\sin(n-m)\varphi\right\}\nonumber\\ 
&=& \varepsilon v^2 \pi\left[
\left(\delta_{m,n+1} + \delta_{m,n-1}\right)\int dr r 
H' + \left(\delta_{m,n-1} - \delta_{m,n+1}\right)\int dr n HK\right],\nonumber
\end{eqnarray}    
i.e. the vortices move under an external force described by   (\ref{Lagrang-interact-s})
only if $m = n \pm 1$. On the same way
one can see that the projection of (\ref{force-2}) onto zero modes along 
$y$-axis is trivial:
\begin{eqnarray}
&&\varepsilon v  \int d^2 x  \varepsilon_{ab}
D_y\phi^a n^b\nonumber\\ 
&=& \varepsilon v^2 \int d^2 x 
\left\{H'\sin\varphi \cos(n-m)\varphi -
\frac{n}{r}HK\cos\varphi\sin(n-m)\varphi\right\}
=0
\end{eqnarray}
for any values of $m,n$.

In order to estimate the integrals let us consider a limiting case.
In the Bogomol'nyi limit 
the first order equations on the shape functions are simply \cite{Bogom}
$$
H' = \frac{n}{r}HK; \qquad \frac{2n}{r}K' = m^2(H^2-1)
$$
Thus the force acting on the vortex in this limit goes to  
\begin{equation}  
{\cal F}_x \longrightarrow 2\pi 
\varepsilon n v^2 \delta_{m,n-1}\int dr HK; \qquad 
{\cal F}_y = 0.
\end{equation}
Thus, the effect of interaction is nontrivial only if $m=n-1$. This is just the
situation when the perturbation term, 
being considered as a correction to the Higgs potential,
lifts the degeneration of the vacuum\footnote{This corresponds to the so-called  thin-wall
approximation of the problem of spontaneous vacuum decay \cite{Vol}.}.  
For example we can consider a simple case when a vortex 
of a unitary magnetic flux interacts with a trivial external 
homogenious perturbation. Note, that then we again have the situation
when the force is orthogonal to the external perturbation. 

Furter aproximation is to neglect the core structure of the vortex, 
i.e. suppose, as it was done at the end of the previous section,  
that everywhere $H \sim 1, K \approx \delta(0)$. 
Then the force acts on 
the configuration in the London limit and can be approximated as 
$$
{\cal F}_x \longrightarrow \pi n \varepsilon v^2 =
\frac{\varepsilon m_v^2}{e} \Phi_n; \qquad 
{\cal F}_y = 0,
$$
that gives the acceleration of the vortex 
$$
W = \frac{\varepsilon m_v^2}{Me} \Phi_n .$$ 
This result can be compared with the final formula from the previous section 
(\ref{accel}). In both situations the force is orthogonal to the perturbation
and the acceleration is proportional to the 
magnetic flux of the vortex and $m_v^2$. 

\subsection*{Conclusions}
This work is based on a paper with V.~Kiselev \cite{KisSh}. 
We have found a regular perturbation scheme to describe the motion of vortices
in the Abelian Higgs model under external force. The key point of our approach is to treat
the excitation of the zero modes of the vortex as a nontrivial time-dependent
translation of the whole configuration. The amplitude of this excitation can be
calculated from the field equation. 
We have considered two different kinds of external 
perturbation connected with current-current interaction and with an 
external scalar
field. In both cases the force acting on the vortex is orthogonal to the 
external perturbation.  

\bigskip I would like to thank Valera Kiselev for very helpful
discussions and many years of fruitful collaboration. I am very
grateful to D.~Maison and V.I.~Zacharov for hospitality at the Max Planck Institut 
f\"ur Physik, Werner-Heisenberg Institut, Munich, where this work was completed.


\begin{thebibliography}{abcdefgh} 
\bibitem{NielsOles} 
H.B. Nielsen and P. Olesen,  Nucl.\ Phys.,\ {\bf B61} (1973) 45. 
\bibitem{Abrikos}
A.A.Abrikosov, Zh.\ Eksp.\  Teor.\ Fiz.,\ {\bf 32} (1957) 1442.
\bibitem{Kibble}T.W.B.~Kibble, J.\ Phys.,\ {\bf A9} (1976) 1387. 
\bibitem{KKS}B.~Kleihaus, J.~Kunz and Ya.~Shnir, Phys.\ Rev.\  {\bf
    D68} (2003) 101701.  
\bibitem{Weinberg} E.J. Weinberg, Phys.\ Rev.\  {\bf D19} (1979) 3008.
\bibitem{Taubes}C. Taubes, Comm.\ Math.\ Phys.,\ {\bf 72} (1980) 277. 
\bibitem{Ball}M. Baker, J.S. Ball and F. Zachariasen, Phys.\ Rev.,\ {\bf D34} (1986) 188.
\bibitem{Rubak}P.J. Rubak, Nucl.\ Phys.,\ {\bf B296} (1988) 669.
\bibitem{Samols}T.M. Samols, Comm.\ Math.\ Phys.,\ {\bf 145} (1992) 149.
\bibitem{Shell}E.P.S.~Shellard and P.J.~Ruback,  Phys.\ Lett.,\ {\bf B209} 262.
\bibitem{Betten}L.M.A. Bettencourt and R.J. Rivers,  Phys.\ Rev.\  {\bf D51} (1995) 1842.
\bibitem{KisSh}V.~Kiselev and Ya.~Shnir, Phys.\ Rev.\  {\bf D57},  
(1998) 5174.
\bibitem{Manton-77} 
N.S. Manton,  Nucl.\ Phys.\ {\bf B126} (1977) 525. 
\bibitem{Rossi} P.~Rossi,  Phys.\ Rep. {\bf 86}, 317 (1982).
\bibitem{Radj}R. Rajaraman, Solitons and Instantons in Quantum Field Theory 
(North-Holland Publ. Co., Amsterdam, 1992). 
\bibitem{Sp}J.M. Speight, Phys. Rev., {\bf D55} (1997) 3830. 
\bibitem{Bogom}E.B.~Bogomolny, Sov.\ J.\ Nucl.\ Phys.\ {\bf 24} (1976) 861;
\bibitem{Vol} M.B.~Voloshin, I.Yu.~Kobzarev and L.B.~Okun, Sov.\ J.\ Nucl.\ Phys.,\ 
{\bf 20}, 644 (1975). 
\end{thebibliography}
\end{document}